\documentclass{amsart}
\usepackage{amsmath}
\usepackage{amssymb}
\input epsf
\newcommand{\si}[1]{\sigma_{#1}}

\newcommand{\W}[4]{\begin{cases}
#1 ,&#2\\
#3 ,&#4
\end{cases}}
\newcommand{\ro}{\rho}
\newcommand{\la}{\lambda}
\newcommand{\La}{\Lambda}

\newcommand{\I}{\mathbb I}

\newcommand{\cH}{{\mathcal H}}
\newcommand{\C}{\mathbb C}
\newcommand{\R}{\mathbb R}
\newcommand{\fA}{\mathfrak A}
\newcommand{\fE}{\mathcal E}
\newcommand{\tr}{\mathrm{tr}\,}
\newcommand{\ptr}[1]{\mathrm{tr}_{#1}}
\newcommand{\tl}[1]{\boldsymbol #1}

\newtheorem{tw}{Theorem}[section]

\newtheorem{uwaga}{Remark}[section]
\begin{document}
\begin{center}
\begin{LARGE}
\textbf{Bell inequalities versus entanglement and mixedness\\[6mm]
for a class of two - qubit states}
\end{LARGE}\\[12mm]
{\L}. Derkacz and L. Jak{\'o}bczyk\\[2mm]
Institute of Theoretical Physics\\ University of
Wroc{\l}aw\\
Pl. M. Borna 9, 50-204 Wroc{\l}aw, Poland
\end{center}
\vskip 8mm \noindent {\sc Abstract}: For a class of mixed two -
qubit states we show that it is not possible to discriminate
between states violating or non - violating Bell - CHSH
inequalities, knowing only their entanglement and mixedness. For a
large set of possible values of these quantities, we construct
pairs of states with the same entanglement and mixedness such
that one state is violating but the other is non - violating Bell
- CHSH inequality.
\section{Introduction}
\noindent
Contradiction between quantum theory and local realism manifests
by the violation of Bell - CHSH inequalities \cite{Bell, CHSH}. It
is  experimental fact that those inequalities can indeed be
violated in many quantum systems (for review see e.g. \cite{Aspect}). On the
other hand, it is well known that quantum states violating
Bell inequalities have to be entangled \cite{E, S}. Since all pure
entangled states violate Bell inequalities \cite{Gisin}, it was
believed that entanglement is equivalent to such violation. After
the work of Werner \cite{Werner}, it turned out that violation of
Bell inequalities is not neccesary for mixed states entanglement.
Thus the relation between entanglement and Bell inequalities is
not clear and  is the interesting problem that should be
investigated in details.
\par
In the case of two - qubit system, there is an effective criterion
for violating the CHSH inequalities \cite{HHH,H}. It enables to
associate with any two - qubit state some numerical parameter
ranging from $0$ for "local states" to $1$ for states maximally
violating such inequalities. Using this criterion, one can study
for example the relation between entanglement, the CHSH violation
and their behaviour under the local filtering operations
\cite{Ver}. Another interesing question is the following: what is
the connection between entanglement and mixedness of the state,
and the amount of CHSH violation given by that state. It is known
that to produce an equal amount of CHSH violation some states
require more entanglement then others. In Ref. \cite{MNW}, it was
suggested that if the more mixed is a state, the higher degree of
entanglement is required for it to violate CHSH inequality.
However there are examples of states that counter that suggestion.
One can find states with equal  amount of CHSH violation and
entanglement, but one of them is more mixed that other. Moreover,
one can construct such states that for fixed CHSH violation, the
order of mixedness for them is always reserved with respect to
the order of their entanglements \cite{Ghosh}.
\par
In the present paper, we study another aspects of the relationship
between entanglement, measured by concurrence $C(\ro)$, mixedness
measured by linear entropy $S_{L}(\ro)$ and CHSH violation. We
ask the following question: is it possible to discriminate
between states violating or non - violating  CHSH inequalities
computing only their entanglement and mixedness? We solve the
problem for some class of mixed two - qubit states.  We show that
there is a large set of possible values of entanglement and
mixedness such that for fixed pair $(s,c)$ in that set, we can
always construct states $\ro_{1},\; \ro_{2}$ with
$C(\ro_{1})=C(\ro_{2})=c$ and $S_{L}(\ro_{1})=S_{L}(\ro_{2})=s$,
such that $\ro_{1}$ is violating CHSH inequality but $\ro_{2}$ is
not violating this inequality. On the other hand, there is also a
subset on the $(s,c)$ plane such that the corresponding states
always violate CHSH inequalities, and the other subset to which
correspond non - violating states. Our results indicate that the
reason why given mixed state violates Bell - CHSH inequlity can
not be explained by their entanglement and mixedness alone.
\section{Violation of Bell inequalities for a pair of qubits}
\subsection{Entanglement}
Consider two-level system $A$ (one- qubit) with the Hilbert space
$\cH_{A}=\C^{2}$ and the algebra of observables $\fA_{A}$ given by
$2\times 2$ complex matrices.
For a joint system $AB$
of two qubits $A$ and $B$, the algebra $\fA_{AB}$ is
equal to $4\times 4$ complex matrices and the Hilbert space
$\cH_{AB}=\cH_{A}\otimes \cH_{B}=\C^{4}$. Let $\fE_{AB}$ be the
set of all states of the compound system i.e.
\begin{equation} \fE_{AB}=\{
\ro\in \fA_{AB}\, : \, \ro\geq 0\quad\text{and}\quad\tr \ro =1 \}
\end{equation}
The state $\ro\in \fE_{AB}$ is \textit{separable}
\cite{Werner}, if it has the form
\begin{equation}
\ro=\sum\limits_{k}\la_{k}\ro_{k}^{A}\otimes \ro_{k}^{B},\quad
\ro_{k}^{A}\in \fE_{A},\;\ro_{k}^{B}\in \fE_{B},\; \la_{k}\geq
0\quad\text{and}\quad \sum\limits_{k}\la_{k}=1
\end{equation}
The set
$\fE_{AB}^{\,\rm sep}$ of all separable states forms a convex
subset of $\fE_{AB}$. When $\ro$ is not separable, it is called
\textit{inseparable} or \textit{entangled}. Thus
\begin{equation}
\fE_{AB}^{\,\rm ent}=\fE_{AB}\setminus \fE_{AB}^{\,\rm sep}
\end{equation}
As a measure of the amount of entanglement a given state contains
we take the entanglement of formation \cite{Bennett}
\begin{equation}
E(\ro)=\min \, \sum\limits_{k}\la_{k}E(P_{k})
\end{equation}
where the minimum is taken over all possible decompositions
\begin{equation}
\ro=\sum\limits_{k}\la_{k}P_{k}
\end{equation}
and
\begin{equation}
E(P)=-\tr[ (\ptr{A}P)\,\log_{2}\, (\ptr{A}P)]
\end{equation}
In the case of two qubits, $E(\ro)$  is the function of another useful
quantity $C(\ro)$ called \textit{concurrence}, which also can be
taken as a measure of entanglement \cite{HW, W}. $C(\ro)$ is
defined as follows
\begin{equation} C(\ro)=\max\;
(\,0, 2p_{\mathrm{max}}(\widehat{\ro})-\tr \widehat{\ro}\,)
\end{equation}
where $p_{\mathrm{max}}(\widehat{\ro})$ denotes the maximal
eigenvalue of $\widehat{\ro}$ and
\begin{equation}
\widehat{\ro}=(\ro^{1/2}\ro^{\dag}\ro^{1/2})^{1/2}
\end{equation}
with
\begin{equation}
\ro^{\dag}=(\si{2}\otimes
\si{2})\,\overline{\ro}\,(\si{2}\otimes \si{2})
\end{equation}
The value of the number $C(\ro)$ varies from $0$ for separable
states, to $1$ for maximally entangled pure states. For the
class  $\fE_{0}$ of states consisting of density matrices of the
form
\begin{equation}
\ro=\begin{pmatrix}0&0&0&0\\
0&\ro_{22}&\ro_{23}&0\\
0&\ro_{32}&\ro_{33}&0\\
0&0&0&\ro_{44}
\end{pmatrix}
\end{equation}
$C(\ro)$ is given by
\begin{equation}
C(\ro)=|\ro_{23}|+\sqrt{\ro_{22}\ro_{33}}-|\,|\ro_{23}|-\sqrt{\ro_{22}\ro_{33}}\,|
\end{equation}
By positive-definiteness of $\ro$, $|\ro_{23}|\leq
\sqrt{\ro_{22}\ro_{33}}$, thus
\begin{equation}
C(\ro)=2\,|\ro_{23}|
\end{equation}
\subsection{Bell - CHSH inequalities}
Let $\tl{a},\,\tl{a}^{\prime},\,\tl{b},\,\tl{b}^{\prime}$ be the
unit vectors in $\R^{3}$ and
$\tl{\sigma}=(\si{1},\,\si{2},\,\si{3})$. Consider the family of
operators on $\cH_{AB}$
\begin{equation}
B_{CHSH}=\tl{a}\cdot\tl{\sigma}\otimes
(\tl{b}+\tl{b}^{\prime})\cdot\tl{\sigma}+\tl{a}^{\prime}\cdot\tl{\sigma}\otimes
(\tl{b}-\tl{b}^{\prime})\cdot\tl{\sigma}
\end{equation}
Then Bell - CHSH  \cite{CHSH} inequalities are
\begin{equation}
|\tr (\ro\, B_{CHSH})|\leq 2
\end{equation}
If the above inequality is not satisfied by  the state $\ro$ for
some choice of $\tl{a},\,\tl{a}^{\prime},\, \tl{b},\,
\tl{b}^{\prime}$ , we say that $\ro$ \textit{violates Bell
inequalities} ($\ro$ \textit{is VBI}). In the case of two-qubit system,
the violation of Bell - CHSH inequalities by mixed states can be
studied using simple necessary and sufficient condition
\cite{HHH,H}.  Any state $\ro\in\fE_{AB}$ can be written as
\begin{equation}
\ro=\frac{1}{4}\left(\I_{2}\otimes\I_{2}+\tl{r}\cdot\tl{\sigma}\otimes\I_{2}+\I_{2}\otimes
\tl{s}\cdot
\tl{\sigma}+\sum\limits_{n,m=1}^{3}t_{nm}\,\si{n}\otimes\si{m}\right)
\end{equation}
where $\I_{2}$ is the identity matrix in two dimensions,
$\tl{r},\tl{s}$ are vectors in $\R^{3}$ and
$\tl{r}\cdot\tl{\sigma}=\sum\limits_{j=1}^{3}r_{j}\si{j}$. The
coefficients
\begin{equation}
t_{nm}=\tr (\ro\,\si{n}\otimes\si{m})
\end{equation}
form a real matrix $T_{\ro}$. Define also real symmetric matrix
\begin{equation}
U_{\ro}=T_{\ro}^{T}\,T_{\ro}
\end{equation}
where $T_{\ro}^{T}$ is the transposition of $T_{\ro}$. Violation
of inequality (14) by the density matrix (15)  and some Bell
operator (13) can be checked by the following criterion: Let
\begin{equation}
m(\ro)=\max_{j<k}\; (u_{j}+u_{k})
\end{equation}
and $u_{j},\, j=1,2,3$ are the eigenvalues of $U_{\ro}$. As was
shown in \cite{HHH,H}
\begin{equation}
\max_{B_{CHSH}}\,\tr (\ro\, B_{CHSH})=2\,\sqrt{m(\ro)}
\end{equation}
Thus (14) is violated by some choice of
$\tl{a},\tl{a}^{\prime},\tl{b},\tl{b}^{\prime}$ iff $m(\ro)>1$.
We can also introduce another parameter
$$
n(\ro)= \max (0,\, m(\ro)-1\,)
$$
ranging from $0$ for non VBI states to $1$ for state maximally
VBI. For the class $\fE_{0}$ we obtain the following expression
for $m(\ro)$
\begin{equation}
m(\ro)=\max\; (2\,C^{2}(\ro),\,(1-2\,\ro_{44})^{2}+C^{2}(\ro)\,)
\end{equation}
where $C(\ro) $ is the concurrence of the state $\ro$. Notice
that all states $\ro\in \fE_{0}$ with concurrence greater then
$\frac{1}{\sqrt{2}}$ are VBI. In the next section we focus on
states with $C(\ro)\leq \frac{1}{\sqrt{2}}$.
\section{The main result}
\noindent Consider now the relation between mixedness,
entanglement and violation of Bell inequalities for mixed states
from the class $\fE_{0}$. Since for $C(\ro)> \frac{1}{\sqrt{2}}$
every mixed state is VBI, consider $\ro$ such that $C(\ro)\leq
\frac{1}{\sqrt{2}}$. Then $m(\ro)>1$ when
\begin{equation}
(1-2\ro_{44})^{2}+C^{2}(\ro)>1
\end{equation}
The above inequality is equivalent to
$$
|\ro_{23}|^{2}>\ro_{44}(1-\ro_{44})
$$
Let us introduce the normalized linear entropy of the state $\ro$
$$
S_{L}(\ro)=\frac{4}{3}\,(1-\tr \ro^{2})
$$
as a measure of its mixedness. We see that $S_{L}(\ro)=0$ for pure
states and $S_{L}(\frac{1}{4}\I_{4})=1$. For  states from the
class $\fE_{0}$
$$
S_{L}(\ro)=\frac{8}{3}\,(\ro_{22}\ro_{33}+\ro_{22}\ro_{44}+\ro_{33}\ro_{44}-|\ro_{23}|^{2})
$$
On the other hand
$$
|\ro_{23}|^{2}-\ro_{44}\,(\ro_{22}+\ro_{33}\,)=|\ro_{23}|^{2}-\ro_{44}\,(1-\ro_{44}\,)>0
$$
so
$$
\ro_{22}\ro_{33}-\frac{3}{8}\,S_{L}(\ro)=|\ro_{23}|^{2}-\ro_{44}
\,(\ro_{22}+\ro_{33}\,)>0
$$
Thus inequality (21) is satisfied iff \cite{JaJam}
\begin{equation}
\ro_{22}\ro_{33}>\frac{3}{8}\, S_{L}(\ro)
\end{equation}
Inequality (22) indicates that states $\ro$ with sufficiently
small mixedness and non - zero entanglement should be VBI. On the
other hand, large mixedness should lead to non - violation of any
Bell inequality. Below we show that there is also another
possibility. For the intermediate values of mixedness, there
exist states with the same linear entropy and concurrence and
such that one of them is VBI, but the other is not VBI. To study
this problem, introduce the subset $\La_{\fE_{0}}\subset \R^{2}$
\begin{equation}
\La_{\fE_{0}}=\{ (\,S_{L}(\ro),\,C(\ro)\,)\,:\,
C(\ro)>0\quad\text{and}\quad \ro\in \fE_{0}\}
\end{equation}
\begin{tw}
\begin{equation}
\La_{\fE_{0}}=\{ (s,c)\in \R^{2}\,:\, 0<c\leq 1,\; 0\leq s\leq
S_{max}(c)\}
\end{equation}
where
$$
S_{max}(C)=\W{\frac{ 8}{ 9}-\frac{ 2}{ 3}\,c^{2}}{c<\frac{ 2}{
3}}{\frac{ 8}{ 3}\, c(1-c)}{c\geq \frac{2}{ 3}}
$$
\end{tw}
\noindent \textit{Proof:} We parametrize the  states $\ro\in
\fE_{0}$ as follows
\begin{equation}
\ro=\begin{pmatrix} 0&0&0&0\\
0&a&\frac{1}{2}c\,e^{i\theta}&0\\
0&\frac{1}{2}c\,e^{-i\theta}&b&0\\
0&0&0&1-a-b
\end{pmatrix},\quad a,b\geq 0,\, \theta\in [0,2\pi]
\end{equation}
Then positive definiteness of $\ro$ is equivalent to
\begin{equation}
ab\geq \frac{c^{2}}{4}\quad\text{and}\quad a+b\leq 1
\end{equation}
On the other hand,
\begin{equation}
S_{L}(\ro)=\frac{4}{3}\,
\left(1-a^{2}-b^{2}-(1-(a+b))^{2}-\frac{c^{2}}{2}\right)
\end{equation}
We are looking for maximal value of (27) for fixed $c$ and $a,b$
such that conditions (26) are satisfied. It turns out that for
$c\in \left(0,\frac{2}{3}\right)$, maximal value of $S_{L}$ is
attained at $a= b=\frac{1}{3}$ and is given by
\begin{equation}
S_{max}(c)=\frac{8}{9}-\frac{2}{3}c^{2},\quad c\in
\left(0,\frac{2}{3}\right)
\end{equation}
For $c\in \left[\frac{2}{3},1\right],\; S_{max}(c)$ is attained at
$a=b=\frac{c}{2}$, thus
\begin{equation}
S_{max}(c)=\frac{8}{3}\, c(1-c),\quad c\in
\left[\frac{2}{3},1\right]
\end{equation}
\begin{uwaga}
Notice that $S_{max}(c)$ is realized by states
\begin{equation}
\ro_{1}(c)=\begin{pmatrix}
0&0&0&0\\0&\frac{1}{3}&\frac{1}{2}c\,e^{i\theta}&0\\
0&\frac{1}{2}c\,e^{-i\theta}&\frac{1}{3}&0\\
0&0&0&\frac{1}{3}
\end{pmatrix},\quad c\in \left(0,\frac{2}{3}\right)
\end{equation}
end
\begin{equation}
\ro_{2}(c)=\begin{pmatrix}
0&0&0&0\\0&\frac{c}{2}&\frac{1}{2}c\,e^{i\theta}&0\\
0&\frac{1}{2}c\,e^{-i\theta}&\frac{c}{2}&0\\
0&0&0&1-c
\end{pmatrix},\quad c\in \left[\frac{2}{3},1\right)
\end{equation}
The states (30) and (31) are locally equivalent to maximally
entangled mixed states discovered in \cite{Munro}. We have obtained
the same result starting from different class of states.
\end{uwaga}
\vskip 2mm\noindent Now consider the structure of the set
$\La_{\fE_{0}}$.
\begin{tw}
$\La_{\fE_{0}}$ is a sum of disjoint subsets $\La_{1},\, \La_{2}$
and $\La_{3}$ with the  properties:
\begin{enumerate}
\item[\textbf{1.}] If $(s,c)\in \La_{1}$, then every state $\ro\in
\fE_{0}$ such that $S_{L}(\ro)=s$ and $C(\ro)=c$ is VBI. \vskip
2mm\noindent
\item[ \textbf{2.}] If $(s,c)\in
\La_{2}$, then there exist states $\ro_{1},\, \ro_{2}\in \fE_{0}$
such that
$$
S_{L}(\ro_{1})=S_{L}(\ro_{2})=s,\quad C(\ro_{1})=C(\ro_{2})=c
$$
and $\ro_{1}$ is VBI,  but $\ro_{2}$ is not VBI. \vskip
2mm\noindent
\item[\textbf{3.}] If $(s,c)\in \La_{3}$, then every state $\ro\in
\fE_{0}$ such that $S_{L}(\ro)=s$ and $C(\ro)=c$ is not VBI.
\end{enumerate}
The sets $\La_{1},\, \La_{2}$ and $\La_{3}$ can be described as
follows (\textbf{Fig. 1}):
\begin{equation*}
\begin{split}
\La_{1}&=\{ (s,c)\,:\, 0<c\leq \frac{1}{\sqrt{2}},\; 0\leq s<
S_{1}(c)\} \cup \{(s,c)\,:\,
\frac{1}{\sqrt{2}}<c\leq 1,\; 0\leq s\leq S_{max}(c)\}\\[2mm]
\La_{2}&=\{ (s,c)\,:\, 0<c\leq \frac{1}{\sqrt{2}},\; S_{1}(c)\leq
s <S_{2}(c)\}\\[2mm]
\La_{3}&=\{ (s,c)\,:\, 0<c\leq \frac{1}{\sqrt{2}},\; S_{2}(c)\leq
s\leq S_{max}(c)\}
\end{split}
\end{equation*}
with
$$
S_{1}(c)=\frac{2}{3}\, c^{2},\quad
S_{2}(c)=\frac{2-c^{2}+2\sqrt{1-c^{2}}}{6}
$$
\end{tw}
\vskip 2mm\noindent
\begin{picture}(300,300)
\put(70,70){\begin{picture}(180,180) \epsffile{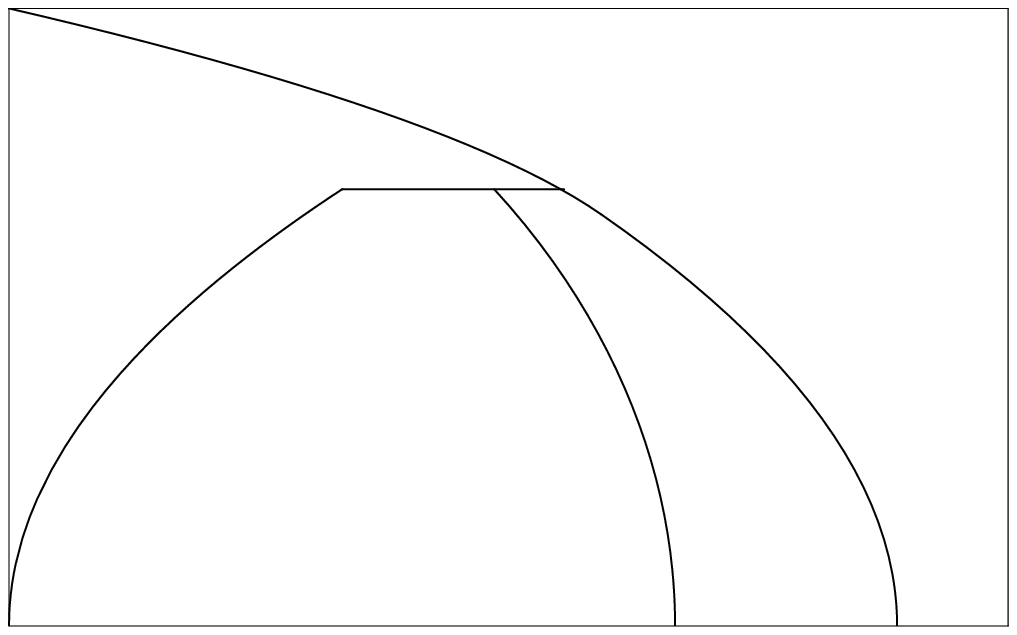}
\end{picture}}
\put(370,65){$S_{L}(\ro)$} \put(45,245){$C(\ro)$}
\put(90,170){$\La_{1}$}
\put(170,100){$\La_{2}$}\put(290,100){$\La_{3}$}
\end{picture}
\vskip -18mm\noindent\centerline{\textbf{Fig. 1.} The set
$\La_{\fE_{0}}$ of admissible pairs $(S_{L}(\ro),\,C(\ro))$ for
$\ro\in \fE_{0}$}
\vskip 8mm \noindent
\textit{Proof:} Consider
the parametrization (25) and introduce new variables
$$
x=\frac{1}{\sqrt{2}}\,(a-b),\quad
y=\frac{1}{\sqrt{2}}\,\left(a+b-\frac{2}{3}\right)
$$
Then conditions (26) can be rewritten as
\begin{equation}
\frac{y^{2}}{2}+\frac{\sqrt{2}y}{3}-\frac{x^{2}}{2}-\frac{c^{2}}{4}+\frac{1}{9}\geq
0
\end{equation}
and
\begin{equation}
y\leq \frac{1}{3\sqrt{2}}
\end{equation}
Thus every point $(x,y)\in X_{+}$, where
$$
X_{+}=\{ (x,y)\,:\,
\frac{y^{2}}{2}+\frac{\sqrt{2}y}{3}-\frac{x^{2}}{2}-\frac{c^{2}}{4}+\frac{1}{9}\geq
0,\;
y\leq \frac{1}{3\sqrt{2}}\}
$$
defines the state $\ro\in \fE_{0}$.
We see
also that
\begin{equation}
S_{L}=-\frac{8}{3}\,\left(\frac{x^{2}}{2}+\frac{3y^{2}}{2}+\frac{c^{2}}{2}-\frac{1}{3}\right)
\end{equation}
and the level set $S_{L}=s$ is the ellipse
\begin{equation}
\frac{x^{2}}{A^{2}}+\frac{y^{2}}{B^{2}}=1
\end{equation}
with
$$
A=\sqrt{6D},\; B=\sqrt{2D},\quad\text{and}\quad
D=-\frac{c^{2}}{12}-\frac{s}{8}+\frac{1}{9}
$$
Thus the set of states with fixed concurrence $C(\ro)=c$ and
linear entropy $S_{L}(\ro)=s$ is determined by the intersection
of the ellipse (35) and $X_{+}$. On the other hand, the condition
(21) equivalent to $m(\ro)>1$ now reads
\begin{equation}
8y^{2}+\frac{4\sqrt{2}}{3}y+c^{2}-\frac{8}{9}>0
\end{equation}
The above inequality can be satisfied by admissible variables $y$ only when
\begin{equation}
y>y_{+}=\frac{-1+3\sqrt{1-c^{2}}}{6\sqrt{2}}
\end{equation}
Similarly, $m(\ro)\leq 1$ for $y\leq y_{+}$. Now the idea of the
proof is simple. For fixed concurrence $c$, the intersection of
the level set of the function $S_{L}$ with $X_{+}$ can lie below
or above the line $y=y_{+}$ or can intersect this line, depending
on the value of $s$ (\textbf{Fig. 2}). The ellipse (35) can
intersect the line $y=y_{+}$ when $B>y_{+}$, thus for
\begin{equation}
s< \frac{2-c^{2}+2\sqrt{1-c^{2}}}{6}
\end{equation}
there are VBI states. The part of ellipse above the line
$y=y_{+}$ represents VBI states, whereas the remaining part
corresponds to states with the same $c$ and $s$, which are not
VBI. For
\begin{equation}
s\geq \frac{2-c^{2}+2\sqrt{1-c^{2}}}{6}
\end{equation}
all states are not VBI.  In the case when the ellipse (35)
intersects hyperbola (32) above the line $y=y_{+}$, all states
are VBI. This can be achieved when
\begin{equation}
s<\frac{2}{3}\,c^{2}
\end{equation}
\vskip 2mm\noindent
\begin{picture}(300,300)
\put(70,70){\begin{picture}(180,180) \epsffile{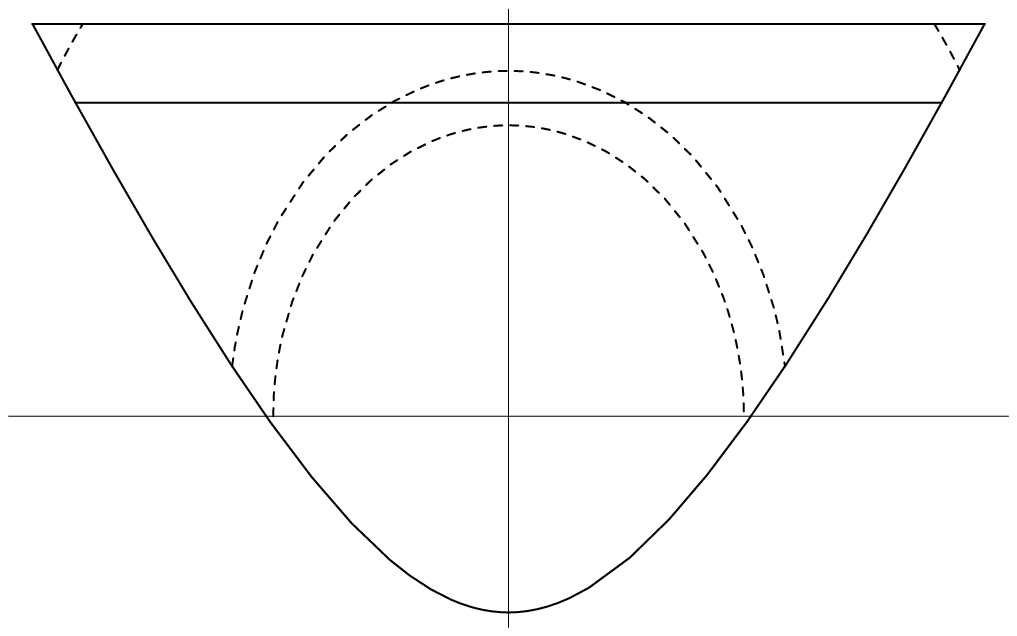}
\end{picture}}\put(366,129){$x$}
\put(212,254){$y$} \put(347,220){$y=y_{+}$} 
\end{picture}
\vskip -18mm \noindent \centerline{\textbf{Fig. 2.} Intersections
of $X_{+}$ with  level sets $S_{L}=s$ (dotted lines) for different
values of $s$.  }
\\[8mm]\noindent
\section{Examples}
\noindent
We can use  parametrization of the ellipse (35) to
construct examples of states with properties listed in Theorem
3.2. If
$$
x=A\cos \varphi,\quad y=B\sin \varphi
$$
then
\begin{equation}
\ro(\varphi,\theta)=\begin{pmatrix}
0&0&0&0\\[2mm]
0&\frac{1}{3}+\sqrt{D}(\sin\varphi
+\sqrt{3}\cos\varphi)&\frac{c}{2}e^{i\theta}&0\\[2mm]
0&\frac{c}{2}e^{-i\theta}&\frac{1}{3}+\sqrt{D}(\sin\varphi
-\sqrt{3}\cos\varphi)&0\\[2mm]
0&0&0&\frac{1}{3}-2\sqrt{D}\sin\varphi
\end{pmatrix}
\end{equation}
where $\theta\in [0,2\pi]$ and $\varphi\in I_{+}$ ($I_{+}$ will depend
on specific values of $c$ and $s$), defines two parameter family
of states with fixed concurrence and linear entropy. The set
$I_{+}$ is defined as follows. Let $c\leq \frac{1}{\sqrt{2}}$. Then:\\[2mm]
\textbf{a.} for $\{(s,c)\,:\,  0\leq s<S_{1}(c)\}\cup
(\La_{2}\cap \{(s,c)\,:\, s< \frac{2}{3}(1-c^{2})\})$
$$
I_{+}=[\varphi_{1},\varphi_{2}]\cup
[\pi-\varphi_{2},\pi-\varphi_{1}]
$$
where
$$
\varphi_{1}=\arcsin
\left[\frac{1}{\sqrt{D}}\,\left(\frac{1}{4}\sqrt{1-3s/2}-1/12\right)\right],\quad
\varphi_{2}=\arcsin\frac{1}{6\sqrt{D}}
$$
\vskip 2mm \noindent \textbf{b.} for $(\La_{2}\cap \{(s,c)\,:\,
s\geq \frac{2}{3}(1-c^{2})\})\cup (\La_{3}\cap\left\{(s,c)\,:\, s\in
[0,\frac{2}{3}],\,c\leq\frac{1}{2}\sqrt{2+2\sqrt{1-\frac{3}{2}s}-\frac{3}{2}s}\right\})$
$$
I_{+}=[\varphi_{1},\pi-\varphi_{1}]
$$
\vskip 2mm\noindent \textbf{c.} for $\La_{3}\setminus\left\{(s,c)\,:\, s\in
[0,\frac{2}{3}],\,c\leq\frac{1}{2}\sqrt{2+2\sqrt{1-\frac{3}{2}s}-\frac{3}{2}s}\right\}$
$$
I_{+}=(0,2\pi]
$$
Define also
$$
\varphi_{3}=\arcsin \left[
\frac{1}{\sqrt{D}}\left(\frac{1}{4}\sqrt{1-c^{2}}-\frac{1}{12}\right)\right]
$$
If $\varphi>\varphi_{3}$, then the points on the ellipse (35)
corresponding to $\varphi$ lie above the line $y=y_{+}$. Thus if
$$
\varphi\in I_{+}\cap I_{B}\quad\text{where}\quad I_{B}=
(\varphi_{3},\pi-\varphi_{3})
$$
all states (41) with such $\varphi$ are VBI. On the other hand, if
$$
\varphi\in I_{+}\setminus I_{B}
$$
all states (41) with such $\varphi$ are not VBI.
So we have:\\[4 mm]
\textbf{1.} If $(s,c)\in \La_{1}$ then $ \varphi_{3}<\varphi_{1}$, and
$$
I_{+}\cap I_{B}=I_{+}\quad\text{and}\quad I_{+}\setminus
I_{B}=\emptyset
$$
\hspace*{4mm} so every state (41) with $\varphi\in I_{+}$  is VBI.\\[2mm]
\textbf{2.} If $(s,c)\in \La_{2}$, both sets $I_{+}\cap I_{B}$
and $I_{+}\setminus I_{B}$ are nonempty. Thus the states (41)
with $\varphi\in I_{+}\cap I_{B}$\\
\hspace*{4mm} are VBI, whereas states with
$\varphi\in I_{+}\setminus I_{B}$  are not VBI.\\[2mm]
\textbf{3.} If $(s,c)\in \La_{3}$, then $\varphi_{3}$ is not
defined and $I_{B}=\emptyset$, so every state (41) with
$\varphi\in I_{+}$ is not VBI.
\newpage\noindent
Consider now the concrete example. Let $c=\frac{1}{2}$ and take
the points
$$
\left(\frac{1}{8},\;\frac{1}{2}\right)\in \La_{1},\;
\left(\frac{1}{2},\; \frac{1}{2}\right)\in \La_{2}\quad
\text{and}\quad \left(\frac{7}{10},\; \frac{1}{2}\right)\in
\La_{3}
$$
Using the parametrization (41), we obtain three families of states
(for simplicity we put $\theta =0$) with corresponding value of
$C(\ro)$ and $S_{L}(\ro)$. So for $s=\frac{1}{8}$ we have the
family
\vskip 8mm\noindent
\begin{equation}
\ro_{1}(\varphi)=\begin{pmatrix}
0&0&0&0\\[2mm]
0&\frac{1}{3}+\frac{\sqrt{43}}{24}\,(\sin\varphi
+\sqrt{3}\cos\varphi)&\frac{1}{4}&0\\[2mm]
0&\frac{1}{4}&\frac{1}{3}+\frac{\sqrt{43}}{24}\,(\sin\varphi
-\sqrt{3}\cos\varphi)&0\\[2mm]
0&0&0&\frac{1}{3}-\frac{\sqrt{43}}{12}\sin\varphi
\end{pmatrix}
\end{equation}
\vskip 8mm \noindent
   with $\varphi\in (0.54657,\,0.65605)$. Then
$m(\ro_{1}(\varphi))>1$ (\textbf{Fig. 3}).\\[8mm]
\begin{picture}(300,300)
\put(70,70){\begin{picture}(180,180) \epsffile{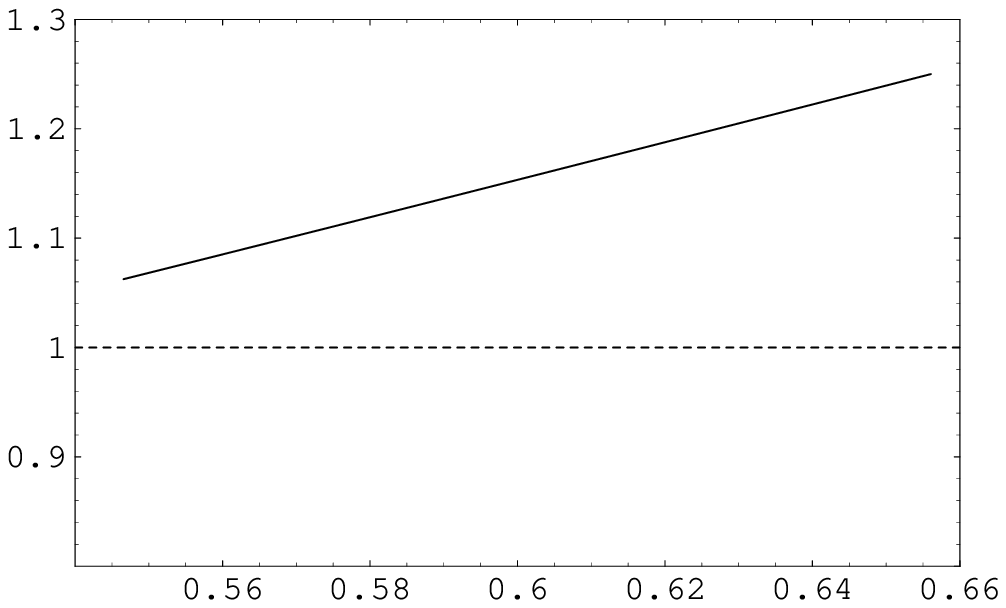}
\end{picture}}\put(370,82){$\varphi$}\put(75,255){$m(\ro)$}
\end{picture}
\vskip -18mm \centerline{\textbf{Fig. 3.} $m(\ro)$ as the function
of $\varphi$ for the states (42)}
\newpage \noindent
Similarly,
for $s=\frac{1}{2}$ we have the family
\begin{equation}
\begin{pmatrix}
0&0&0&0\\[2mm]
0&\frac{1}{3}+\frac{1}{6}\,(\sin\varphi
+\sqrt{3}\cos\varphi)&\frac{1}{4}&0\\[2mm]
0&\frac{1}{4}&\frac{1}{3}+\frac{1}{6}\,(\sin\varphi
-\sqrt{3}\cos\varphi)&0\\[2mm]
0&0&0&\frac{1}{3}-\frac{1}{3}\,\sin\varphi
\end{pmatrix}
\end{equation}
with $\varphi\in (0.25,1.57)$. In that case $m(\ro)$ can be
smaller or bigger then $1$, depending on $\varphi$ (\textbf{Fig.
4}).\\
\begin{picture}(300,300)
\put(70,70){\begin{picture}(180,180) \epsffile{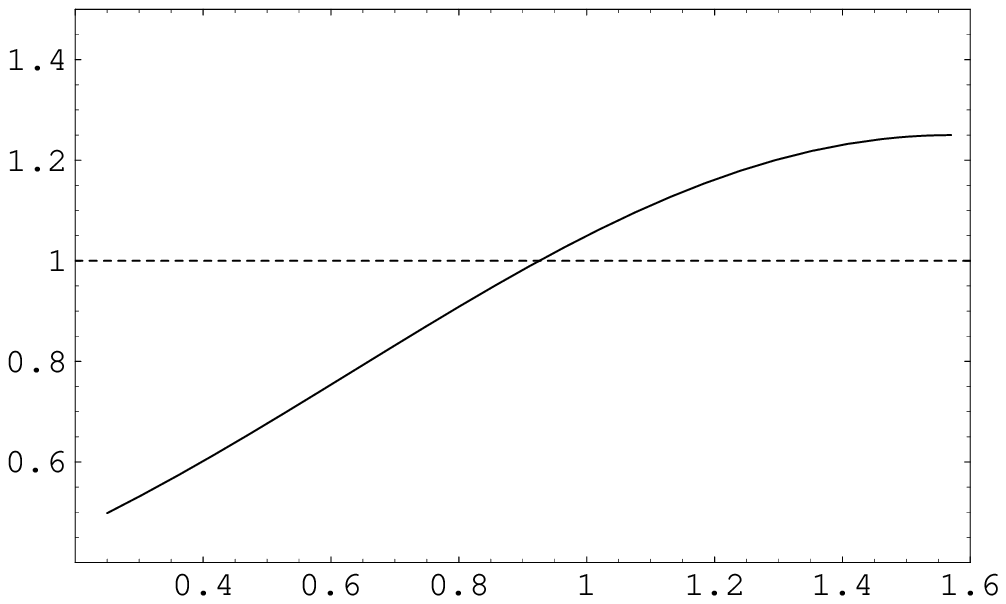}
\end{picture}}\put(370,82){$\varphi$}\put(75,255){$m(\ro)$}
\end{picture}
\vskip -18mm \centerline{\textbf{Fig. 4.} $m(\ro)$ as the function
of $\varphi$ for the states (43)}
\vskip 4mm
\noindent
Finally, for $s=\frac{7}{10}$ we obtain
\vskip 4mm
\noindent
\begin{equation}
\begin{pmatrix}
0&0&0&0\\[2mm]
0&\frac{1}{3}+\frac{1}{6\sqrt{10}}\,(\sin\varphi
+\sqrt{3}\cos\varphi)&\frac{1}{4}&0\\[2mm]
0&\frac{1}{4}&\frac{1}{3}+\frac{1}{6\sqrt{10}}\,(\sin\varphi
-\sqrt{3}\cos\varphi)&0\\[2mm]
0&0&0&\frac{1}{3}-\frac{1}{3\sqrt{10}}\,\sin\varphi
\end{pmatrix}
\end{equation}
with $\varphi \in (0,2\pi)$. For this family $m(\ro)<1$
(\textbf{Fig. 5}).
\newpage\noindent
\begin{picture}(300,300)
\put(70,70){\begin{picture}(180,180) \epsffile{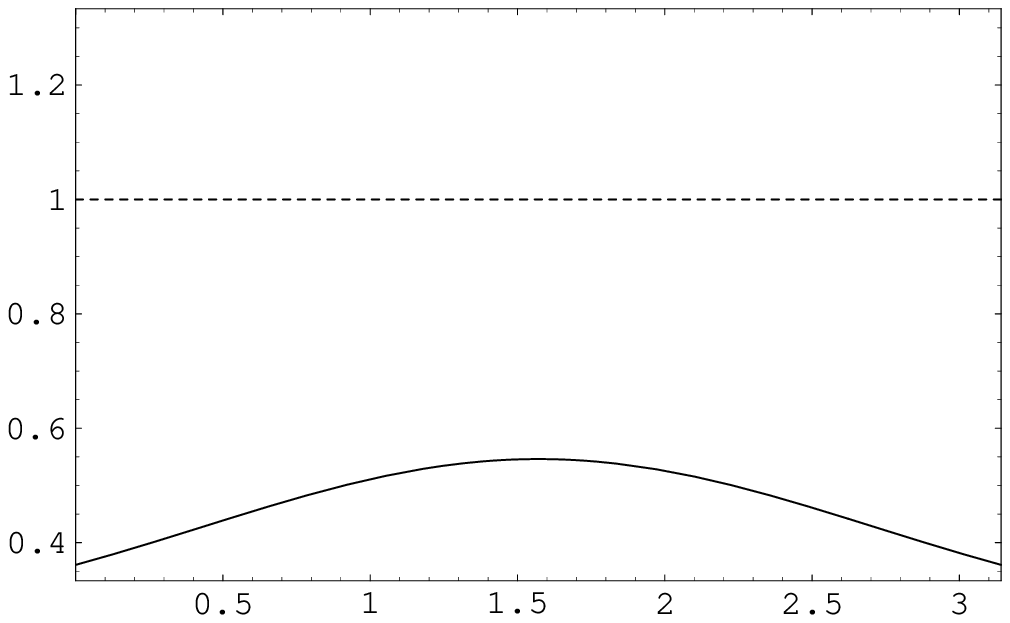}
\end{picture}}\put(370,82){$\varphi$}\put(75,255){$m(\ro)$}
\end{picture}
\vskip -18mm \centerline{\textbf{Fig. 5.} $m(\ro)$ as the function
of $\varphi$ for the states (44)}
\vskip 1cm\noindent
\textbf{Acknowledgments}\\[4mm]
L.J. acknowledges financial support by Polish Ministry of Scientific Research and
Information Technology under the grant
PBZ-Min-008/PO3/2003.

\end{document}